\documentclass[twocolumn]{aastex631}

\usepackage{amsmath}
\usepackage{soul}
\usepackage{natbib}
\usepackage{CJK}
\begin{document}
\begin{CJK*}{UTF8}{gbsn}

\title{A Magnetar Engine and Circumstellar Medium Interaction: Synergistic Effects in Producing Superluminous Supernovae}

\author[0000-0001-8744-3813]{Guang-Lei Wu (吴光磊)}

\author[0000-0002-1067-1911]{Yun-Wei Yu (俞云伟)}

\author[0000-0002-8708-0597]{Liang-Duan Liu (刘良端)}
\affiliation{Institute of Astrophysics, Central China Normal University, Wuhan 430079, China; \url{yuyw@ccnu.edu.cn}}
\affiliation{Laboratory for Compact Object Astrophysics and Astronomical Technology, Central China Normal University, Wuhan 430079, China}
\affiliation{Education Research and Application Center, National Astronomical Data Center, Wuhan 430079, China}

\begin{abstract}
Superluminous supernovae (SLSNe) are often modeled as being powered either by a central engine or by strong interaction with dense circumstellar material (CSM). These two mechanisms may be dynamically coupled if the ejecta interact with dense CSM while being energized by a newborn magnetar. We develop a semi-analytical hybrid model that follows the coupled dynamics, energy conversion, and radiative output of such systems. A rapidly rotating magnetar injects energy through a relativistic wind, inflating a hot bubble inside the expanding ejecta. Part of the injected energy is stored as radiation, while the rest is converted into bulk kinetic energy of the swept-up ejecta. At the same time, the outer ejecta collide with the surrounding CSM and form a circumstellar interaction (CSI) region. As the shock driven by the magnetar accelerates through the ejecta, it can catch up with the CSI region and take over the subsequent interaction with the unshocked CSM.
The emergent light curves are therefore governed by the coupled effects of magnetar energy injection, shock heating, and radiative diffusion. We show that this hybrid model can produce diverse SLSN light-curve morphologies, including luminous interaction-powered peaks, asymmetric post-peak declines, and late-time emission sustained by delayed leakage of magnetar-powered radiation. The model provides a plausible way to reduce the extreme nickel-mass or initial explosion-energy requirements often encountered in purely radioactive or purely interaction-powered interpretations.
\end{abstract}


\section{Introduction} \label{sec:intro}
Superluminous supernovae (SLSNe) are a rare class of supernovae (SNe) characterized by their unusually high luminosity with a peak absolute magnitude of $M<-21$ \citep{Gal-Yam2012, Gal-Yam2019}, in contrast to typical core-collapse SNe (CCSNe) which usually peak at absolute magnitudes of $\sim-15$ to $-19$   \citep{Gal-Yam2017}. In addition to their high luminosity, most SLSNe exhibit a hot blue continuum in their early-stage spectra, corresponding to a high photospheric temperature. If these SLSNe are powered by the radioactive decays of $^{56}$Ni as usual \citep{Arnett1982}, then an exceptionally large mass ($\sim10M_{\odot}$) of radioactive nickel would be required, which is difficult to synthesize in normal CCSNe \citep{Umeda2008}. In principle, pair-instability SNe (PISNe) could synthesize such a large amount of nickel \citep{Barkat1967, Rakavy1967, Heger2002, Kasen2011}, and may provide an explanation for a handful of SLSNe \citep[e.g., SN 2007bi and SN 2018ibb;][]{Gal-Yam2009, Schulze2024}. However, several observational features of most SLSNe could still challenge the PISN model, i.e., (i) the predicted PISN light curves evolve too slowly to match most observed light curves, (ii) the post-peak decline of SLSNe usually cannot be described by an exponential as expected for the radioactive decay of $^{56}$Ni, and (iii) the extremely low metallicity required for PISN host galaxies \citep[$\sim 0.1 Z_{\odot}$;][]{Yusof2013} is usually disfavored by observations \citep{Leloudas2015, Perley2016, Chen2017, Schulze2018, Cleland2023}.

SLSNe can be classified into hydrogen-poor SLSNe~I
and hydrogen-rich SLSNe~II based on whether hydrogen is detected in their spectra \citep{Smith2007, Quimby2011, Gal-Yam2012, Gal-Yam2019}. For some SLSNe~II, the narrow Balmer line appearing in their spectra strongly suggests that they occur in a dense hydrogen-rich circumstellar
medium (CSM) and are powered by the interaction of the SN ejecta with the CSM \citep{Smith2007, Smith2008, Pessi2025}. In this
circumstellar interaction (CSI) scenario, the collision between the SN ejecta and the CSM drives shocks that convert kinetic energy
into thermal energy \citep{Chevalier2011, Ginzburg2012, Chatzopoulos2013, Moriya2013b, Dessart2015, Inserra2018, Khatami2024}.
As the shocks propagate, the thermal energy deposited in the post-shock region gradually 
diffuses outward and is finally released as SN emission. 
This mechanism has been widely employed to explain SNe~II with narrow emission lines (SNe~IIn), suggesting that SLSNe~II could just be the most luminous of this population \citep{Howell2017, Pessi2025}.  Although CSI is most naturally associated with hydrogen-rich events that show clear narrow-line signatures,  it has also been proposed as a possible power source for some SLSNe~I. In such cases, the CSM is expected to be hydrogen-poor, consisting instead of helium- or carbon/oxygen-rich material lost by the progenitor shortly before explosion. Such a scenario may account for the high luminosities of SLSNe~I and can also help explain sharp light-curve peaks or undulations \citep{Chatzopoulos2013, Sorokina2016, Wheeler2017}.

However, the most extreme events still pose a serious challenge to CSI interpretations. Some SLSNe radiate more than several $10^{51}\mathrm{erg}$, exceeding the total explosion energy of normal CCSNe. Meanwhile, the photospheric velocities of SLSNe are found to be higher than those of normal CCSNe, and models combining CSI with $^{56}$Ni decay generally require relatively massive progenitors, implying high kinetic energies for SLSNe \citep{Reka2021, Chen2023a, Chen2023b, Konyves2025}.  Although the inferred zero-age main-sequence masses ($M_{\mathrm{ZAMS}}$) of some SLSNe fall within the range expected for pulsational PISNe (PPISNe), the required initial kinetic energies for some of these SLSNe still exceed the upper limits predicted for PPISNe \citep{Woosley2017}.
Therefore, it seems unavoidable to invoke an unusually powerful central engine to explain SLSNe, even if the observed emission is powered by CSI.

Such a powerful engine has been widely proposed as a millisecond magnetar \citep{Wheeler2000, Kasen2010, Woosley2010, Inserra2013, Metzger2015}. High magnetic fields of up to $\sim10^{15}$ G can arise from the extremely high differential rotation of the proto-neutron star left by the SN explosion \citep{Duncan1992, Thompson1993}. The magnetar model is further supported by numerical simulations (e.g., \citealt{Chen2016, Blondin2017, Suzuki2017, Suzuki2019, Chen2020, Suzuki2021}) and by successful fits to the light curves of SLSNe, particularly SLSNe~I (e.g., \citealt{Liu2017, Nicholl2017, Yu2017, Hsu2021, Chen2023b, Omand2024}).   
To be specific, the rotational energy of a magnetar can be released in the form of a relativistic electron-positron wind. The collision of this wind with the SN ejecta can drive a pair of shocks, including a forward shock propagating in the ejecta and a reverse shock in the wind to create a wind nebula \citep{Kotera2013, Metzger2014, Yu2019b, Yu2019a, Vurm2021}. Therefore, on the one hand, the forward shock could produce a pre-peak bump when it breaks out of the ejecta \citep{Kasen2016, Liu2021, Chen2026} as observed in some SLSNe. On the other hand, the emission from the pulsar wind nebula (PWN) can be absorbed by the ejecta, converting the wind energy to the thermal energy of the ejecta. The gradual leakage of this PWN emission could provide a natural explanation for the observations of oxygen lines in later-time spectra and possible gamma-ray emission \citep{Murase2015, Lunnan2016, Jerkstrand2017, Dessart2019, Murase2021, Omand2023, Li2024, Omand2026}. 

In traditional considerations, the magnetar engine has been invoked only for powering SN emission directly. However, as mentioned above, even in the CSI model, the extremely high energy of the SN ejecta could also ultimately be provided by the central engine. Therefore, in a unified picture, except for a small number of SLSNe attributable to PISNe, a significant fraction of SLSNe could be driven by powerful central engines. The observational diversity of SLSNe may arise from differences in their environments, as first suggested in \citet{Yu2017}. In the literature, some authors have indeed proposed hybrid energy source models to explain some specific SLSNe \cite[e.g.,][]{Chatzopoulos2016, Chen2017b, Inserra2017, Li2020}. In these models, however, engine powering and CSI are considered completely separately and independently, which is likely an oversimplification.
 In this paper, therefore, we present a semi-analytical model to approximately follow the coupled dynamics and emission of the SLSNe in which an inner magnetar energizes ejecta interacting with the surrounding CSM (Section~\ref{sect:Mod}). In Section~\ref{Sec: RA}, we present our numerical results and analytical scalings for the multi-stage dynamical evolution, energy budget, and radiation. We also illustrate that the hybrid model can broadly reproduce the overall morphology of observed light curves and explore the parameter space. Finally, in Section~\ref{Sec: Sum}, we summarize our results and discuss their implications.

\begin{figure*}[htbp]
    \centering
    \includegraphics[width=1\linewidth]{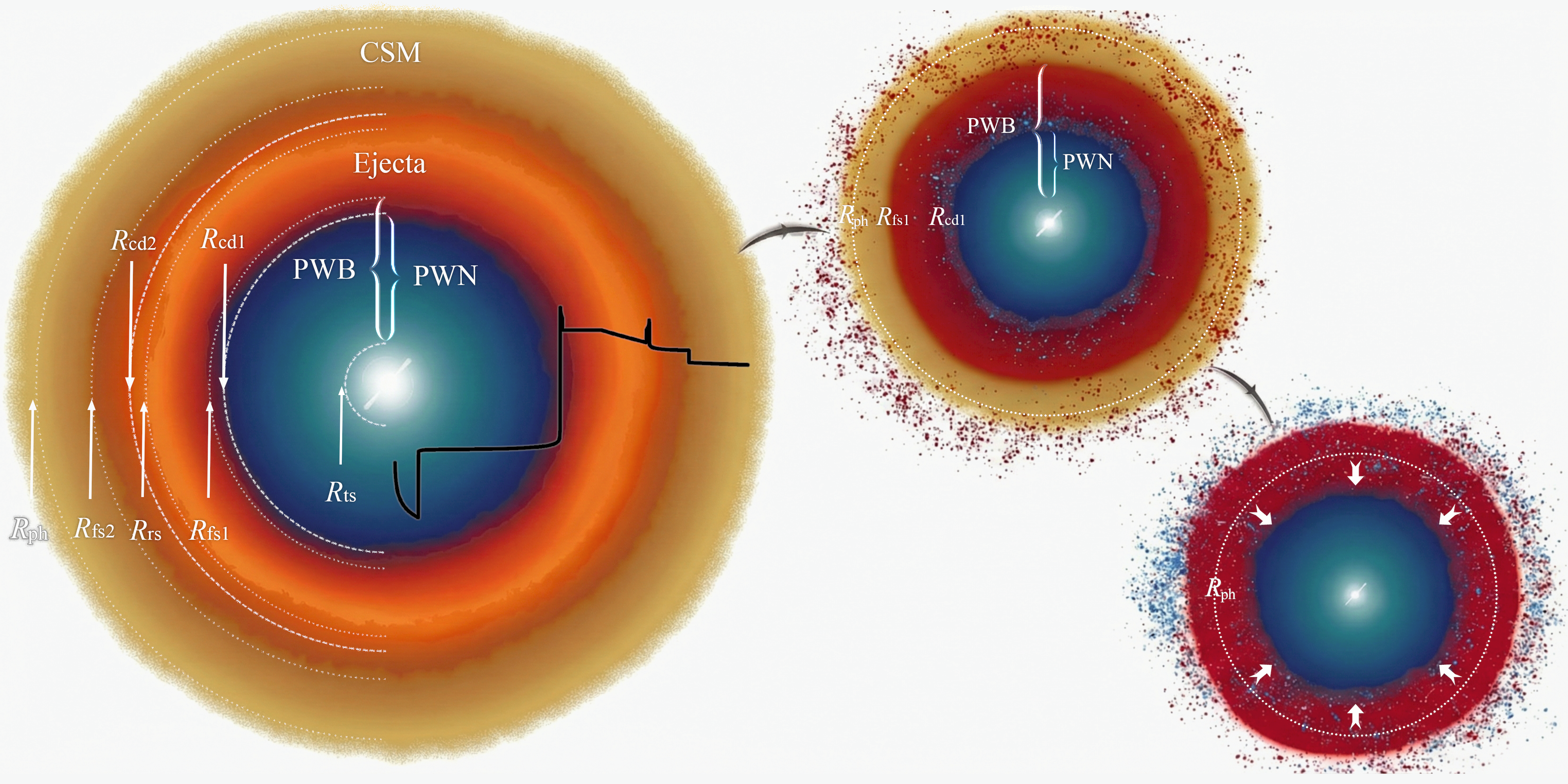}
    \caption{Schematic illustration of the dynamical evolution of our hybrid model,
    from the early coexistence of the PWB and CSI region to the later breakout of FS1 into the CSM.
    The black solid line shows the density profiles inspired by \citet{Chevalier1982} and \citet{Jun1998}.
    The color rendering is assisted by Nano Banana Pro, based on an author-drawn draft.
    }
    \label{ill}
\end{figure*}
\section{The Model}
\label{sect:Mod}
\subsection{Physical inputs}\label{sec:input}
Following the central-engine interpretation commonly invoked for SLSNe, we adopt a rapidly rotating magnetar whose rotational energy is released via magnetic-dipole spin-down. In this case, the spin-down luminosity of a magnetar engine is written as
\begin{equation}
    L_{\rm sd}(t)=\frac{E_{\mathrm{m}}}{t_{\rm sd}}\left(1+{\frac{t}{t_{\rm sd}}} \right)^{-2},
\end{equation}
where $E_{\mathrm{m}}\simeq 2 \times 10^{52}P_{\mathrm{i,-3}}^{-2}~\mathrm{erg}$ is the total rotational energy of the magnetar, $t_{\rm sd}\simeq2\times 10^{5} B_{14}^{-2}P_{\mathrm{i,-3}}^2 ~\mathrm{s}$ is the spin-down timescale, and $B$ and $P_{\mathrm{i}}$ are the surface strength of the dipolar magnetic field and the initial spin period, respectively. Meanwhile, the cold SN ejecta with a mass $M_{\mathrm{ej}}$ is assumed to follow a broken power-law density profile, consisting of a flat inner core and a steep outer envelope \citep{Chevalier1989, Matzner1999}:
\begin{equation}
    \rho_{\mathrm{ej}}(r,t)=\zeta_{\rho}\frac{M_{\mathrm{ej}}}{v_{\mathrm{t}}^3t^3}\begin{cases}\displaystyle\left(\frac{r}{v_\mathrm{t}t}\right)^{-\delta},&r<v_{\mathrm{t}} t,
    \\ \displaystyle\left(\frac{r}{v_{\mathrm{t}}t}\right)^{-n},&  r\geq v_{\mathrm{t}} t,
    \end{cases}
\end{equation}
where $\zeta_{\rho}=(n-3)(3-\delta)/[4\pi(n-\delta)]$. The transition velocity is defined as $v_{\mathrm{t}}=\zeta_{v}\left({E_{\mathrm{sn}}}/{M_{\mathrm{ej}}}\right)^{1/2}$ with $\zeta_v=\sqrt{2(5-\delta)(n-5)/[(n-3)(3-\delta)]}$, and $E_{\mathrm{sn}}$ is the initial kinetic energy of the SN ejecta. Typical values of $\delta$ and $ n $ are usually found in the ranges of $0\leq\delta\leq1$ and $7\leq n\leq12$, and we adopt $\delta=0$ and $n=10$ as fiducial values in our calculations.

The CSM distribution around an SN is usually closely related to the mass loss of its progenitor star during the late stage of evolution. This mass loss may occur via stellar winds \citep{Langer2012, Smith2014} or pulsational pair instability (PPI)-driven mass ejections \citep{Woosley2007, Woosley2017}. We adopt a power-law density profile for the CSM:
\begin{equation}
 \rho_{\mathrm{csm}}\left(r\right)= \rho_{\mathrm{0}}\left(\frac{r}{R_{\mathrm{csm}}}\right)^{-s}   ,
\end{equation}
where $\rho_{\mathrm{0}} \simeq {(3-s)M_{\mathrm{csm}}}/({4\pi R_{\mathrm{csm}}^{3}})$ normalizes the profile to the total CSM mass ($M_{\mathrm{csm}}$) and the outer radius ($R_{\mathrm{csm}}$).  Following common practice, we adopt $s=2$, which corresponds to a wind-like CSM profile expected from steady mass loss at an approximately constant outflow velocity during the late evolutionary stages of the progenitor \citep{Smith2014}.

\subsection{Shock structures}
 
The dynamical interaction of the magnetar wind, SN ejecta, and surrounding CSM has been extensively studied in the context of SN remnants  \citep[e.g., ][]{Reynolds1984, Blondin2001, Swaluw2001, Gelfand2009, Bandiera2020}, where two pairs of shocks can be driven. To be specific, as illustrated in Figure~\ref{ill}, $R_{\mathrm{ts}}$ and $R_{\mathrm{fs1}}$ label the termination shock (TS) of the magnetar wind and the forward shock (FS1) propagating in the SN ejecta, respectively, while the contact discontinuity surface (CD1) between the two shocked regions is located at $R_{\mathrm{cd1}}$. The region between $R_{\mathrm{ts}}$ and $R_{\mathrm{cd1}}$  is referred to as the PWN, and the entire region between $R_{\mathrm{ts}}$ and  $R_{\mathrm{fs1}}$, including the PWN
and the shocked ejecta swept up by FS1, is referred to as the pulsar wind bubble (PWB). The high pressure within the PWN can compress the shocked region behind FS1 into a thin shell, i.e., $(R_{\mathrm{fs1}}-R_{\mathrm{cd1}})\ll R_{\mathrm{fs1}}$ \citep{Chevalier1992, Jun1998, Chevalier2005}, especially for $t<t_{\rm sd}$. Meanwhile, the collision of the SN ejecta with the CSM drives
another pair of shocks, with $R_{\mathrm{rs}}$, $R_{\mathrm{cd2}}$ and $R_{\mathrm{fs2}}$ representing the radii of the reverse shock (RS), the second contact discontinuity surface (CD2) and the outer forward shock (FS2). 

The primary difference between the situation we consider and previous studies of SN remnants is that the SN ejecta is initially highly optically thick, which traps photons and the high-velocity shock is dominated by radiation pressure. In contrast, a typical SN remnant is usually optically thin and dominated by hot gas with a long cooling timescale. Nevertheless, in both cases, the shock evolves effectively adiabatically and exhibits similar dynamics. Therefore, the shock structures discussed above provide the basic framework for our study and play an important role in determining the SN emission.

Another important difference between this work and the evolution of a PWN expanding in an SN remnant is that the total energy released by pulsars born from normal CCSNe is only on the order of $\sim 2\times 10^{50} P_{\mathrm{i,-2}}^{-2}~\mathrm{erg}$, which is much less than the initial kinetic energy of the SN ejecta. Therefore, FS1 is only slowly accelerated through the inner core of the SN ejecta and encounters the RS after a long time ($\sim \rm kyr$). In this case, the pressure of the RS is comparable to or even greater than that of the PWN. As a consequence, the PWN  will be strongly decelerated and confined to a very small region (i.e., $R_{\mathrm{cd1}}\ll R_{\mathrm{fs2}}$) 
\footnote{ 
After the RS collides with FS1 and before the system relaxes into the final subsonic Sedov-Taylor phase, the PWN  undergoes a sequence of compressions and re-expansions \citep{Swaluw2001}. This stage, characterized by damped oscillatory behavior, is referred to as the reverberation phase \citep{Bucciantini2003, Bandiera2023b, Bandiera2023}. }. 
In contrast, in the SLSN scenario considered here, the magnetar can inject energy comparable to or exceeding the initial SN kinetic energy, i.e., $E_{\mathrm{m}}\ge E_{\mathrm{sn}}$, so FS1 is driven outward much more rapidly. In the parameter space relevant to this work, the pressure of the PWB exceeds that of the CSI region when FS1 encounters the RS (see Section~\ref{sec:para} for a detailed discussion). Consequently, FS1 eventually breaks out of the SN ejecta, merges with FS2, and thereafter dominates the dynamical evolution of the system.

\subsection{The Evolution of the PWB} \label{sec: Dyn PWB}

The PWB is assumed to be isobaric with a uniform pressure $P_{\rm pwb}$. The evolution of the kinetic energy of the PWB can be determined by
\begin{equation}
 \frac{d E_{\mathrm{k,pwb}}}{d t} =4\pi R_{\rm fs1}^2P_{\rm pwb}{\frac{dR_{\rm fs1}}{dt}}+\frac{1}{2}v_{\rm us}^2 \frac{dM_{\mathrm{fs1}}}{dt}-H_{\mathrm{fs1}},\label{eq:Ekfs1}
\end{equation}
where the first term on the right side represents the adiabatic acceleration of the region, the second term is the original kinetic energy of the newly swept-up upstream material with $v_{\rm us}$ being the upstream velocity, and
\begin{equation}
H_{\mathrm{fs1}} =\frac{1}{2}\left(v_{\mathrm{fs1}}-v_{\rm us}\right)^2\frac{d M_{\mathrm{fs1}}}{dt}
\end{equation}
is the rate at which the kinetic energy of FS1 is converted into the internal energy of the shocked material. The PWN is assumed to consist mainly of electrons and positrons, and its kinetic energy is negligible compared with that of the shocked material. Substituting $E_{\mathrm{k,pwb}}={\frac{1}{2}}M_{\mathrm{fs1}}v_{\mathrm{fs1}}^2$ into Eq. (\ref{eq:Ekfs1}) gives
\begin{eqnarray}\label{Eqs:vfs1}
 \frac{d v_{\mathrm{fs1}}}{d t} &=&\frac{1}{M_{\mathrm{fs1}}}\left[
     4\pi R_{\rm fs1}^2P_{\rm pwb}-\left( v_{\mathrm{fs1}}-v_{\rm us} \right)\frac{dM_{\mathrm{fs1}}}{dt}
     \right]\nonumber\\
     &=&\frac{4\pi R_{\rm fs1}^2}{M_{\mathrm{fs1}}}\left[
     P_{\rm pwb}-\rho_{\rm us}\left(v_{\mathrm{fs1}}-v_{\rm us}\right)^2
     \right],
\end{eqnarray}
where the radius and mass of FS1 evolve according to
\begin{equation}
 \frac{d R_{\mathrm{fs1}}}{d t} =v_{\mathrm{fs1}}
\end{equation}
and
\begin{equation}
 \frac{d M_{\mathrm{fs1}}}{d t} =4 \pi R_{\mathrm{fs1}}^2\rho_{\rm us}\left(v_{\mathrm{fs1}}-v_{\rm us}\right).
\end{equation}
The upstream density and velocity ahead of FS1 are written as
\begin{equation}
  \rho_{\rm us}(r) = \begin{cases}
    \rho_{\rm ej}, & r < R_{\rm rs},\\
    \rho_{\rm csi}, & R_{\rm rs} \le r < R_{\rm fs2},\\
    \rho_{\rm csm}, & r \ge R_{\rm fs2},
  \end{cases}
\end{equation}
and
\begin{equation}
  v_{\rm us}(r) = \begin{cases}
    v_{\rm ej}, & r < R_{\rm rs},\\
    v_{\rm csi}, & R_{\rm rs} \le r < R_{\rm fs2},\\
    v_{\rm w}, & r \ge R_{\rm fs2},
  \end{cases}
\end{equation}
where $v_{\rm w}\ll v_{\mathrm{fs1}}$ is the velocity of the wind-like CSM, and the subscripts ``ej'', ``csi'', and ``csm'' denote the
SN ejecta, the CSI region, and the CSM,
respectively. The quantities in the CSI region between the RS and FS2 can be determined by solving the hydrodynamic equations using similarity variables \citep{Chevalier1982}.

The solution of the above dynamical equations is highly dependent on the determination of the pressure $P_{\rm pwb}$. For a radiation-dominated PWB, the pressure is related to the total internal energy by $P_{\rm pwb} = U_{\mathrm{pwb}}/(3V_{\mathrm{pwb}})$, where $V_{\mathrm{pwb}} \approx (4\pi/3) R_{\rm fs1}^3$ for $R_{\rm fs1} \gg R_{\rm ts}$.
The PWB consists of the PWN and the FS1 region with internal energies $U_{\mathrm{pwn}}$ and $U_{\mathrm{fs1}}$, respectively, so $U_{\mathrm{pwb}}=U_{\mathrm{pwn}}+U_{\mathrm{fs1}}$.  
The evolution of the internal energies in these two regions is governed by
\begin{equation}\label{eqs: Upwn}
     \frac{d U_\mathrm{pwn}}{dt} =  L_{\rm sd}- L_{\mathrm{pwn}}-4\pi P_{\rm pwb}R_{\rm cd1}^2v_{\rm cd1}
\end{equation}
and
\begin{equation}
     \frac{d U_\mathrm{fs1}}{dt} =  H_{\rm fs1}- L_{\mathrm{fs1}}-4\pi P_{\rm pwb}(R_{\rm fs1}^2v_{\rm fs1}-R_{\rm cd1}^2v_{\rm cd1}),
\end{equation}
where $R_{\mathrm{cd1}}=(U_{\mathrm{pwn}}/U_{\mathrm{pwb}})^{1/3}R_{\mathrm{fs1}}$ is used, and $L_{\mathrm{fs1}}$ and $L_{\mathrm{pwn}}$ are the emission luminosities from the FS1 region and the PWN, respectively. The last term in Equation~(\ref{eqs: Upwn}) represents the work done by the expanding PWN on the FS1 region across CD1.
According to the photon diffusion in these regions, we adopt
\begin{equation}\label{eqs: Lfs1}
L_{\mathrm{fs1}}\approx\frac{U_{\mathrm{fs1}}}{t_{\rm d,fs1}},
\end{equation}
where $t_{\rm d,fs1}$ is the diffusion timescale from the FS1 region to the photosphere. The photospheric radius $R_{\mathrm{ph}}$ is defined by the equation $\kappa\int_{R_{\mathrm{ph}}}^{\max [R_{\mathrm{csm}},R_{\mathrm{fs1}}]} \rho dr=2/3$ with $\kappa\approx0.2~\mathrm{cm^2~g^{-1}}$  adopted as a commonly used constant optical opacity for ionized H-poor material. Before FS1 reaches the photosphere, $R_{\mathrm{ph}}$ lies at a fixed position in the CSM as discussed in Section~\ref{HAR}. After $R_{\mathrm{fs1}}>R_{\mathrm{ph}}$, the photosphere is determined by the same condition and moves into the FS1 region.
Considering that the density of the FS1 region varies significantly from its innermost to outermost surface, its photon diffusion timescale should be defined by averaging the specific timescales from $R_{\mathrm{cd1}}$ to $R_{\mathrm{fs1}}$ as
\begin{equation}\label{eqs: tdfs1}
t_{\rm d,fs1}
=
\left(\frac{\displaystyle\int_{R_{\rm cd1}}^{R_{\rm fs1}}
t^{-1}_{\rm d}(r)r^2dr}{\displaystyle
 \int_{R_{\rm cd1}}^{R_{\rm fs1}} r^2dr}\right)^{-1},
\end{equation}
where the weight of $4\pi r^2 dr$ is required for the spherical symmetry \footnote{In the diffusion approximation, a thin layer at radius $r$ between $R_{\mathrm{cd1}}$ and $R_{\mathrm{fs1}}$ contributes to the emergent luminosity as $dL\simeq dU/t_{\mathrm{d}}(r)$, where $dU= 3P_{\mathrm{pwb}}\times4\pi r^2dr$ and $U_{\mathrm{fs1}}=\int_{R_\mathrm{cd1}}^{R_{\mathrm{fs1}}} dU $. An effective diffusion timescale  $t_{\mathrm{d,fs1}}$ is then defined through
$L_{\mathrm{fs1}}\approx U_{\mathrm{fs1}}/t_{\mathrm{d,fs1}}=\int_{R_\mathrm{cd1}}^{R_{\mathrm{fs1}}}dL$, which yields Equation~(\ref{eqs: tdfs1}).}. Here, the diffusion timescale as a function of radius $r$ is given by
\citep{Ginzburg2012}
\begin{equation}
\begin{aligned}
        t_{\mathrm d}(r)=&\int^{{R_{\mathrm{ph}}}}_{r} \frac{\kappa\rho(r') d(r'-r)^2}{c}.
\end{aligned}
\end{equation}
Meanwhile, we also have
\begin{equation}
    L_{\mathrm{pwn}}\approx \frac{U_{\mathrm{pwn}}}{t_{\mathrm{d,pwn}}},
\end{equation}
with
\begin{equation}\label{eqs:tdpwn}
\begin{aligned}
        t_{\mathrm {d,pwn}}=&\int^{{R_{\mathrm{ph}}}}_{R_{\mathrm{cd1}}}  \frac{\kappa\rho(r') d(r'-R_{\mathrm{cd1}})^2}{c}+\frac{R_{\mathrm{cd1}}}{c},
\end{aligned}
\end{equation}
where the emission generated by the PWN is considered to be non-thermal and primarily in the high-energy bands. For the baryon-starved PWN, photon propagation is approximated by the light-crossing time $R_{\mathrm{cd1}}/c$ rather than a diffusion time. The high-energy radiation escaping from the PWN is partially absorbed by the FS1 region and the unshocked CSM, where it is then reprocessed into thermal emission. We parametrize the thermalized fraction as
$L_{\rm{pwn,th}}=L_{\rm{pwn}}(1 - e^{-\tau_{\gamma}})$ , where $\tau_{\gamma} \approx \kappa_{\gamma} (M_{\mathrm{fs1}} / 4\pi R_{\mathrm{fs1}}^2+\int_{R_\mathrm{fs1}}^{R_{\mathrm{csm}}}\rho_{\mathrm{us}}dr)$ is the optical depth through the FS1 region and the CSM with $\kappa_{\gamma}$ being an effective gamma-ray opacity for PWN emission. 
This treatment is analogous to the gamma-ray leakage prescription introduced for the radioactive decays of
$^{56}$Ni and $^{56}$Co
\citep{Clocchiatti1997,
Chatzopoulos2012}, but the value of $\kappa_{\gamma}$ can vary over a range of $\sim0.01$--$0.1~\mathrm{cm^2~g^{-1}}$ \citep{Nicholl2017,Gomez2024} since
the thermalization efficiency of PWN
emission instead depends on the nebular spectrum and ejecta ionization state
\citep{Vurm2021}.  
  The thermal emission associated with the PWB is then taken to be the sum of the FS1 emission
and the thermalized contribution from the PWN, i.e., $L_{\rm{pwb}}=L_{\rm{pwn,th}}+L_{\rm{fs1}}$.
In evaluating the above diffusion timescales, we adopt the density profile 
\begin{equation}
    \rho(r)=\begin{cases}
        \rho_{\mathrm{fs1}}, &R_{\mathrm{cd1}}\leq r<R_{\mathrm{fs1}},\\
        \rho_{\mathrm{us}}, &r \geq R_{\mathrm{fs1}},
    \end{cases}    
\end{equation}
where $\rho_{\mathrm{fs1}}$ is taken to be uniform in the FS1 region, with
\begin{equation}
\rho_{\mathrm{fs1}}=\frac{M_{\mathrm{fs1}}}{\displaystyle \frac{4}{3}\pi(R_{\mathrm{fs1}}^3-R_{\mathrm{cd1}}^3)}.
\end{equation}

\subsection{The RS and FS2 dynamics}
In analogy with the PWB case, the evolution of the kinetic energy of the CSI region can be determined by
\begin{equation}\label{eqs:CSM shock kinetic energy}
\begin{aligned}
     \frac{dE_{\mathrm{k,csi}}}{dt}=        &4\pi R_{\mathrm{fs2}}^2P_{\mathrm{fs2}}v_{\mathrm{fs2}}-4\pi R_{\mathrm{rs}}^2P_{\mathrm{rs}}v_{\mathrm{rs}}+\\&
        \frac{1}{2}v_{\mathrm{w}}^2\frac{dM_{\mathrm{fs2}}}{dt}+\frac{1}{2}v_{\mathrm{ej}}^2\frac{dM_{\mathrm{rs}}}{dt}-H_{\mathrm{fs2}}-H_{\mathrm{rs}},
\end{aligned}
\end{equation}
where  
\begin{equation}
     H_{\mathrm{fs2}}=\frac{1}{2}(v_{\mathrm{fs2}}-v_{\mathrm{w}})^2\frac{dM_{\mathrm{fs2}}}{dt}
\end{equation}
and
\begin{equation}
    H_{\mathrm{rs}}=\frac{1}{2}(v_{\mathrm{ej}}-v_{\mathrm{rs}})^2\frac{dM_{\mathrm{rs}}}{dt}
\end{equation}
are the conversion of the bulk kinetic energy of the incoming ejecta and CSM into the thermal energy after the two shocks. 
Here, the swept-up masses in the FS2 and RS regions are given by
\begin{equation}
    \frac{dM_{\mathrm{rs}}}{dt}
    = 4\pi R_{\mathrm{rs}}^2\rho_{\mathrm{ej}}(v_{\mathrm{ej}}-v_{\mathrm{rs}})
\end{equation}
and
\begin{equation}
     \frac{dM_{\mathrm{fs2}}}{dt}
    = 4\pi R_{\mathrm{fs2}}^2\rho_{\mathrm{csm}}(v_{\mathrm{fs2}}-v_{\mathrm{w}}),
\end{equation}
respectively.
Combining the above equations with the expression of $E_{\mathrm{k,csi}}=\tfrac{1}{2}M_{\rm rs}v_{\rm rs}^2+\tfrac{1}{2}M_{\rm fs2}v_{\rm fs2}^2$, we obtain
\begin{eqnarray}
\label{eqs: vfs2}
M_{\mathrm{fs2}}v_{\mathrm{fs2}}\frac{dv_{\mathrm{fs2}}}{dt}
&=&4\pi R_{\mathrm{fs2}}^2\left[P_{\mathrm{fs2}}-\rho_{\mathrm{csm}}(v_{\mathrm{fs2}}-v_{\mathrm{w}})^2\right]v_{\mathrm{fs2}}\nonumber\\
&&-4\pi R_{\mathrm{rs}}^2\left[P_{\mathrm{rs}}-\rho_{\mathrm{ej}}(v_{\mathrm{ej}}-v_{\mathrm{rs}})^2\right]v_{\mathrm{rs}}\nonumber\\
&&-M_{\mathrm{rs}}v_{\mathrm{rs}}\frac{dv_{\mathrm{rs}}}{dt}.
\end{eqnarray}
Here, the quantities in the RS region can be related to those in the FS2 region using the self-similar solution, which is valid when the RS propagates into the steep outer envelope of the ejecta and FS2 moves through the CSM. In this case, the time-independent velocity and pressure profiles within the shocked region can be written as
\begin{eqnarray}
{v(r,t)}=f_v(\eta)v_{\rm fs2}(t),\\
{P(r,t) }=f_P(\eta)P_{\rm fs2}(t),\label{similarP}
\end{eqnarray}
where $\eta \equiv r/R_{\rm fs2}(t)$ is the similarity variable, and the expressions of the dimensionless profile functions $f_v(\eta)$ and $f_P(\eta)$ for different density indices $n$ and $s$ are given by \cite{Chevalier1982}. Using Equation~(\ref{similarP}), the pressure behind FS2 can be further related to the average pressure of the CSI region, which is given by $P_{\mathrm{csi}}=\int_{R_{\mathrm{rs}}}^{R_{\mathrm{fs2}}}r^2P(r,t)dr/\int_{R_{\mathrm{rs}}}^{R_{\mathrm{fs2}}}r^2 dr$. The average pressure can then be determined by $P_{\mathrm{csi}}\approx U_{\mathrm{{csi}}}/(3V_{\mathrm{csi}})$, with $V_{\mathrm{csi}}\approx4\pi(R_{\mathrm{fs2}}^3-R_{\mathrm{rs}}^3)/3$ and
\begin{equation}\label{eqs:Ucsi}
    \frac{dU_{\mathrm{csi}}}{dt}= H_{\mathrm{fs2}}+H_{\mathrm{rs}}-L_{\mathrm{csi}}-(4\pi R_{\mathrm{fs2}}^2P_{\mathrm{fs2}}v_{\mathrm{fs2}}-4\pi R_{\mathrm{rs}}^2P_{\mathrm{rs}}v_{\mathrm{rs}}),
\end{equation}
where 
\begin{equation}\label{eqs: Lfs2}
L_{\mathrm{csi}}\approx\frac{U_\mathrm{csi}}{ t_{\mathrm{d,csi}}}
\end{equation}
is the emission luminosity of the CSI region. In view of the CSI region being a thin shell, its photon diffusion time is shorter than that ahead of FS2, and hence we adopt
\begin{equation}\label{eqs:tdcsi}
t_{\mathrm{d,csi}}\simeq\int_{R_\mathrm{fs2}}^{R_\mathrm{ph}}\frac{\kappa\rho_{\mathrm{csm}}(r')d(r'-R_{\mathrm{fs2}})^2}{c}.
\end{equation}

\section{Results and Analyses}\label{Sec: RA}

\subsection{Dynamics}\label{sec:dyn}

\begin{figure}
    \centering
    \includegraphics[width=1\linewidth]{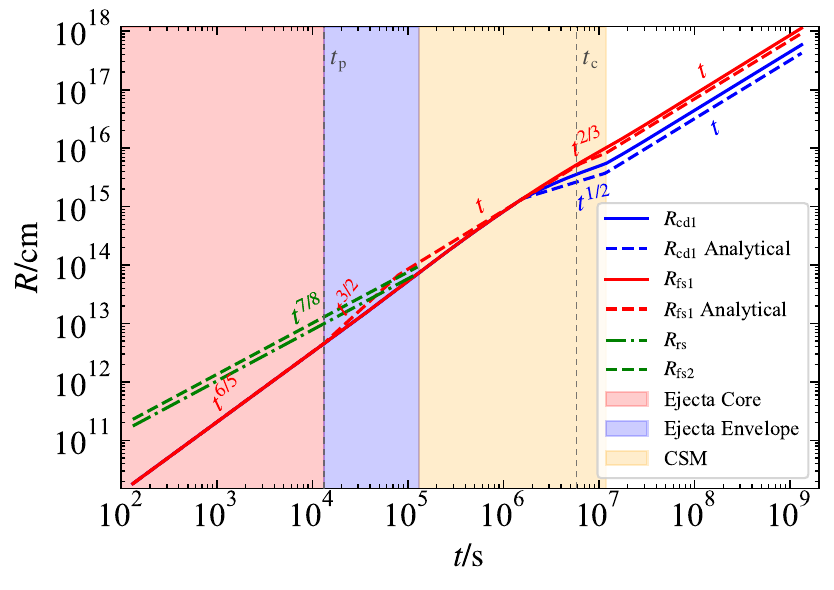}
    \caption{Evolutions of different radii. Different background colors indicate that FS1 is moving in different regions.}
    \label{figvRU}
\end{figure}

For a fiducial set of model parameters as $L_{\rm sd,i}=E_{\rm m}/t_{\rm sd}=10^{47}\mathrm{erg~s^{-1}}$, $t_{\rm sd}=10^5\mathrm{s}$, $E_{\rm sn}=10^{51}\rm erg$, $M_{\mathrm{ej}}=10M_{\mathrm{\odot}}$, $R_{\mathrm{csm}}=10^{16}\rm cm$, and  $M_{\mathrm{csm}}=5M_{\mathrm{\odot}}$, we present the numerical results for the dynamical evolution in Figure~\ref{figvRU}, where some analytical approximations are also presented. The dynamical evolution of the PWB can be roughly divided into four stages. 

First, the PWB expands within the inner core until it reaches the transition radius  $R_{\mathrm{t}}=v_{\mathrm{t}}t$ between the inner core and the outer envelope of the ejecta. Its
dynamics are similar to those of a PWN expanding into an SN remnant and have been widely studied in previous works developed to describe the dynamical evolution of SN remnants \citep[e.g.,][]{Chevalier1977, Chevalier1984, Reynolds1984, Chevalier1992, Jun1998, Chevalier2005}.
More specifically, for this stage with $L_{\rm pwn}\simeq 0$, the solution for the evolution of the PWB can be approximated by \citep{Chevalier2005}
\begin{equation}\label{eqs: Rp}
     R_{\mathrm{cd1}}\simeq R_{\mathrm{fs1}}\simeq v_{\mathrm{t}} t_{\mathrm{p}}\left({\frac{t}{t_{\rm p}}}\right)^{\alpha},
\end{equation}
where $\alpha=(6-\delta)/(5-\delta)$ at this stage,  $t_{\mathrm{p}}=\zeta_{\mathrm{p}} (E_{\mathrm{sn}}/E_{\mathrm{m}})t_{\rm sd}$ is the time at which the PWB reaches the transition radius, and $\zeta_{\mathrm{p}} =2[(n-5)(9-2\delta)(11-2\delta)]/[(5-\delta)^2(n-\delta)(3-\delta)]$.

In stage II,  the PWB expands into the outer envelope of the ejecta. The index $\alpha$ then changes and asymptotically approaches $3/2$, which can be obtained by setting  $1/2 M_{\mathrm{fs1}}R_{\mathrm{fs1}}^2/t^2 \sim L_{\rm sd,i} t$ where $M_{\mathrm{fs1}}$ approaches a constant value as $R_{\mathrm{fs1}}\rightarrow\infty$ \footnote{Recent studies show that forward shock accelerates faster than the bulk of the shocked ejecta in the outer envelope of the SN ejecta \citep{Blondin2017, Suzuki2017, Govreen-Segal2021, Chen2026}.  In this situation, FS1 gradually separates from CD1, and a steep pressure gradient develops between CD1 and FS1, forming a ``blowout layer". Our simplified model does not capture this detailed structure and instead follows only the bulk motion. After FS1 enters the CSM, the shallow power‑law density profile suppresses further acceleration, and the FS1 velocity rapidly relaxes toward the characteristic bulk velocity of the swept‑up material. }.  
For $t > t_{\mathrm{sd}}$, the PWB cannot be further accelerated by the magnetar and therefore approaches free expansion, so that
\begin{equation}\label{eqs:Rp2}
     R_{\mathrm{cd1}}\simeq R_{\mathrm{fs1}} \simeq v_{\mathrm{t}} t_{\mathrm{p}}\left({\frac{t_{\rm sd}}{t_{\rm p}} }\right)^{\alpha}{\frac{t}{ t_{\rm sd}}}.
\end{equation}
At the same time, the RS propagates inward through the outer envelope of the freely expanding ejecta toward the transition radius. 
The dynamics of the RS and FS2 can likewise be characterized by  
\begin{equation}\label{eqs:Rc}
    R_{\mathrm{rs}}\simeq R_{\mathrm{fs2}}\simeq  v_{\mathrm{t}} t_{\mathrm{c}}\left({\frac{t}{t_{\rm c}} }\right)^{\beta},
\end{equation}
where $\beta=(n-3)/(n-s)$. The time at which the RS reaches the transition radius $R_{\mathrm{t}}$ can be expressed as \citep{Chevalier2003, Moriya2013}
\begin{equation}\label{eqs:tc}
     t_{\mathrm{c}}=\zeta_\mathrm{c} \left(\frac{ M_{\mathrm{ej}}}{M_{\mathrm{csm}} }  \right)^{\frac{1}{3-s}}\frac{R_{\mathrm{csm}}}{v_{\mathrm{t}}},
 \end{equation}
where $\zeta_{\mathrm{c}}=[(4-s)(3-\delta)/(n-4)(n-\delta)]^{1/(3-s)}$. For convenience, the characteristic timescales introduced in the model and used in the analysis are summarized in Table~\ref{tab:timescales}.  As the swept-up mass becomes dynamically important, $\beta$ gradually approaches the
Sedov–Taylor value of
$2/(5-s)$. The RS continues to move inward toward the center with an approximately constant velocity in the comoving frame of the ejecta \citep{Truelove1999, Laming2003, Hwang2012}.
As the PWB expands further, FS1 eventually collides with the RS.
Because the mass loading from the CSI region is insufficient to compress the PWB, the PWB undergoes only mild deceleration. 
Subsequently, FS1 breaks out of FS2, marking the transition from stage II to stage III.

In stage III, the PWB  begins to interact with the CSM. At the beginning of this stage, only a small amount of CSM has been swept up, and the PWB continues to expand nearly ballistically. As more mass is accumulated, the PWB gradually relaxes towards the Sedov-Taylor solution, in which the radius of FS1 evolves as $R_{\mathrm{fs1}}\propto t^{2/(5-s)}$ \citep{Sedov1959, Zeldovich1967}.
According to the shock jump conditions, the pressure downstream of the freshly shocked CSM declines as $P_{\mathrm{fs1}}\propto t^{-6/(5-s)}$.  Meanwhile, the pressure in the PWN evolves as $P_{\mathrm{cd1}}\propto R_{\mathrm{cd1}}^{-4}$ and pressure equilibrium implies that $R_{\mathrm{cd1}}\propto t^{3/(10-2s)}$.
Finally, in stage IV, the PWB has traversed the entire CSM and enters a phase of homologous expansion.
 
\begin{deluxetable}{lll}
\tablecaption{Characteristic timescales used in this work.\label{tab:timescales}}
\tablehead{
\multicolumn{1}{l}{Notation} & \multicolumn{1}{l}{Definition} & \multicolumn{1}{l}{Interpretation}
}
\tabletypesize{\scriptsize}
\tablewidth{0pt}
\startdata
$t_{\rm sd}$ & $E_{\rm m}/L_{\rm sd,i}$ & magnetar spin-down timescale \\
$t_{\rm d,fs1}$ & Equation~(\ref{eqs: tdfs1}) & diffusion timescale from FS1 \\
$t_{\rm d,pwn}$ & Equation~(\ref{eqs:tdpwn}) & diffusion timescale from PWN \\
$t_{\rm d,csi}$ & Equation~(\ref{eqs:tdcsi}) & diffusion timescale from CSI \\
$t_{\rm p}$ & $\zeta_{\rm p}(E_{\rm sn}/E_{\rm m})t_{\rm sd}$ & time when the PWB reaches $R_{\rm t}$ \\
$t_{\rm c}$ & Equation~(\ref{eqs:tc}) & time when the RS reaches $R_{\rm t}$ \\
$t_{\rm b}$ & Equation~(\ref{eqs:tb}) & time when FS1 catches up with FS2 \\
$t_{\rm pb}$ & Equation~(\ref{eqs:tpb}) & timescale of the post-peak bump \\
\enddata
\end{deluxetable}

\subsection{Heating and radiation}\label{HAR}

\begin{figure}
    \centering
    \includegraphics[width=1\linewidth]{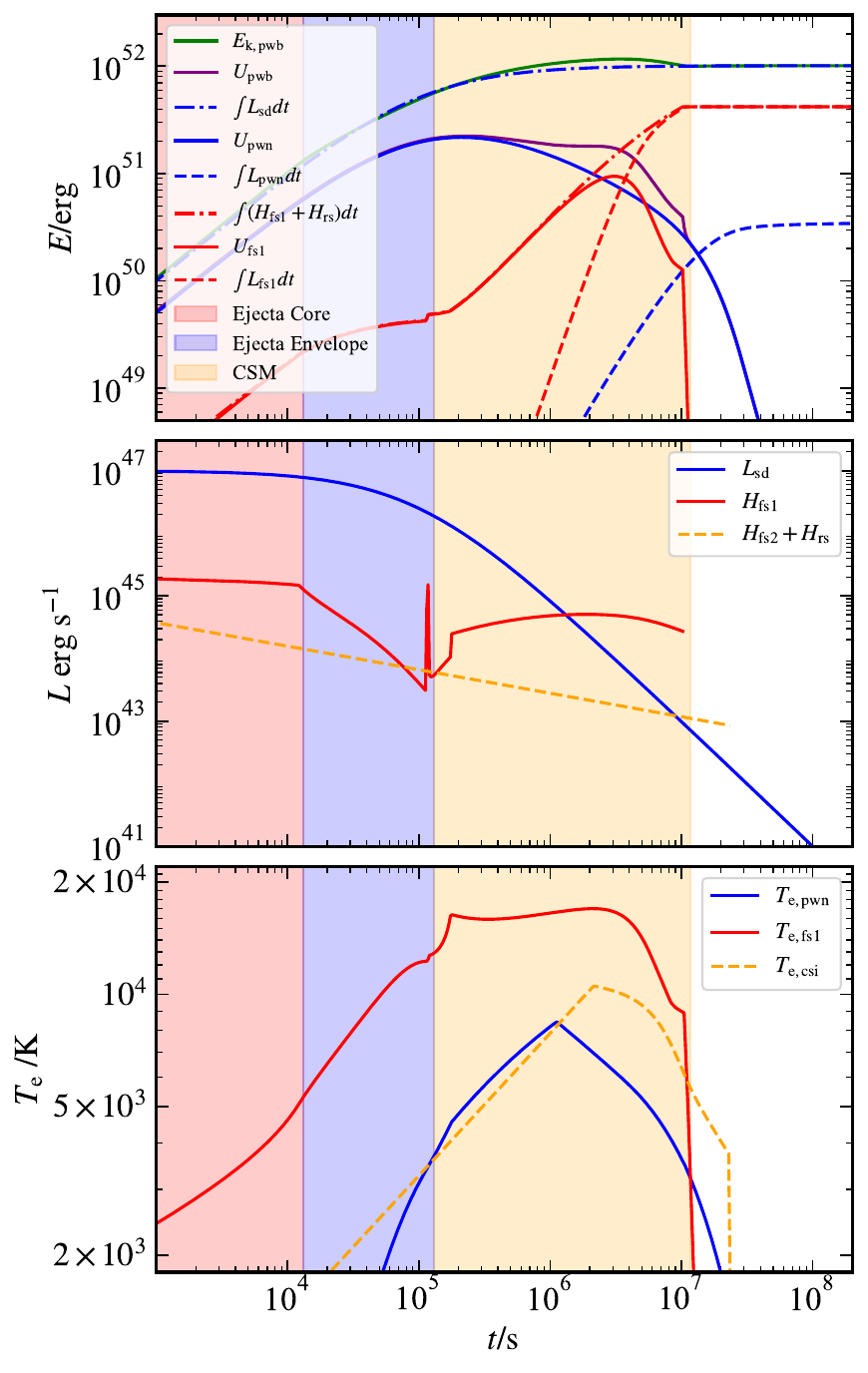}
    \caption{Evolution of the accumulated energies of different components (top),  instantaneous heating rate (middle), and effective emission temperature (bottom). Different background colors indicate the region in which  FS1 is moving. For comparison, the dashed orange line shows the CSI component for the model without a magnetar.}
    \label{heating rate}
\end{figure}
Having outlined the dynamical evolution across the four stages, we now examine how the energy budget and heating evolve during these stages. 
In the first stage, as shown in the top panel of Figure~\ref{heating rate}, a constant fraction of the spin-down energy deposited in the PWN  is stored as internal energy, such that $U_{\mathrm{pwn}}/\int L_{\rm sd}dt =1/(1+\alpha)$. The remainder is converted into the kinetic energy of the PWB through work done across CD1.  Only a small fraction, $U_{\mathrm{fs1}}\simeq \int H_{\mathrm{fs1}}dt=(3-\delta)/[2(11-2\delta)(9-2\delta)]L_{\rm sd,i}t$, is deposited as internal energy between $R_{\mathrm{cd1}}$ and $R_{\mathrm{fs1}}$ \citep{Chevalier2005}.
As a result, the region between $R_{\mathrm{cd1}}$ and $R_{\mathrm{fs1}}$ is compressed into a very thin shell.  

In stage II,  the steep density decline in the outer ejecta leads to a rapid decrease in the shock heating rate $H_{\mathrm{fs1}}$, thereby slowing the growth of the internal energy $U_{\mathrm{fs1}}$.
The presence of CSM substantially modifies the thermalization process. In the absence of CSM, shock heating deposited behind  FS1 can produce early shock breakout emission once FS1 catches up with the receding photosphere \citep{Kasen2016, Liu2021, Chen2026}. As the ejecta become optically thin, energy previously deposited in the PWN is gradually released and powers the SN light curve. In contrast, when CSM is present, FS1 does not break out immediately as the photosphere remains embedded within the CSM at a fixed radius.  Instead, the PWB continues to expand outward until it collides with the CSI region. Because of the high density of the CSI region, this collision produces a sharp peak in the heating rate, as shown in the middle panel of Figure~\ref{heating rate}. As a result, the CSI region is destroyed and no longer contributes to the emission. For comparison, Figure~\ref{heating rate} also shows the CSI component in the absence of the PWB.

In stage III, shock heating becomes important as FS1 interacts with the CSM, converting bulk kinetic energy into internal energy and increasing $U_{\mathrm{fs1}}$. As long as the upstream optical depth satisfies $\tau_{\mathrm{us}}
>c/v_{\mathrm{fs1}}$, FS1 remains radiation mediated and efficiently traps the thermal photons. When the optical depth drops below $\tau_{\mathrm{us}}
<c/v_{\mathrm{fs1}}$, photons begin to escape from FS1 and diffuse through the CSM, enhancing the radiative losses from the PWB and causing  $U_{\mathrm{fs1}}$ and $U_{\mathrm{pwn}}$ to decline. In this semi-analytical treatment, we approximate the emission emerging from FS1 as quasi-blackbody until the upstream optical depth drops below $\tau_{\mathrm{us}}<2/3$ \footnote{This criterion differs somewhat from that of \citet{Chatzopoulos2013}, who assumed that shock heating ceases to be efficiently deposited into the radiation once $\tau<c/v$. Although such shocks are no longer radiation mediated, the shocked gas can still emit thermal bremsstrahlung radiation \citep{Chevalier2012, Svirski2012}, which can be reprocessed by the CSM if it remains effectively optically thick. In this case, the observed spectrum may evolve from an approximately thermal shape to a more complex, transmitted spectrum \citep{Wasserman2025, Govreen-Segal2026}. }. For comparison among the different emission components, we adopt an effective blackbody temperature for each component $i$ as $T_{\mathrm{e},i}=(L_i/4\pi R_{\mathrm{ph}}^2\sigma)^{1/4}$, 
where $L_i$ denotes the luminosity of the corresponding component and $\sigma$ is the Stefan-Boltzmann constant. The resulting temperature evolution is shown in the bottom panel of Figure~\ref{heating rate}.
The temperature evolution in the hybrid model differs from that of the CSI component in the absence of a PWB. The peak effective temperature is mainly set by the FS1 emission and exceeds that of the CSI component. The effective temperature associated with the PWN, which is governed by delayed magnetar-powered heating, remains lower than that of the CSI component.

\begin{figure}
    \centering
    \includegraphics[width=1\linewidth]{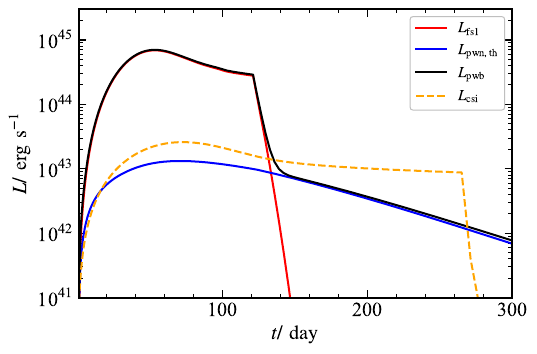}
    \caption{The bolometric luminosities of different components, where the dashed orange line shows the CSI emission in the model without a magnetar.}
    \label{lightcurve}
\end{figure}
\begin{figure*}
    \centering
    \includegraphics[width=1\linewidth]{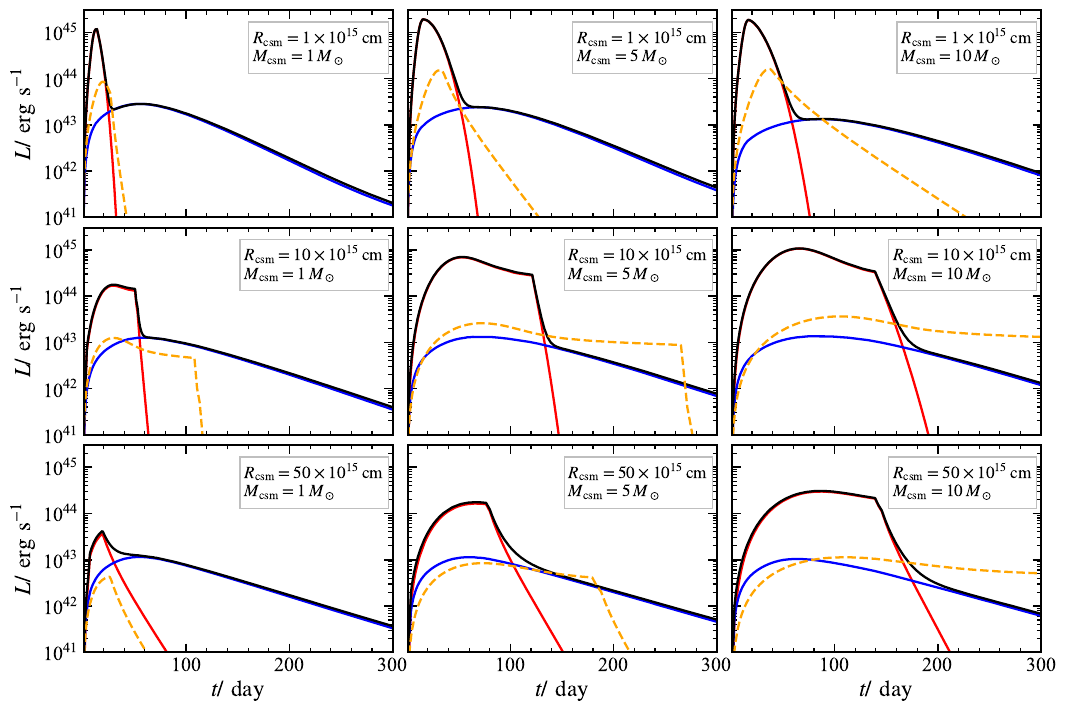}
    \caption{Same as Figure~\ref{lightcurve}, but with different $R_{\mathrm{csm}}$ and $M_{\mathrm{csm}}$.}
    \label{blc}
\end{figure*}

Figure~\ref{lightcurve} presents the resulting bolometric light curves. The FS1 component powers the light-curve peak, after which the emission enters a rapid cooling phase. Under the same SN and CSM parameters, the FS1 emission is more luminous than the CSI emission because of the higher shock velocity. At later times, the light curve is dominated by the PWN emission, which is powered by internal energy deposited by the magnetar and emerges after diffusion through the shocked material. Figure~\ref{blc} further shows the bolometric light curves for different CSM radii and masses. For relatively small $M_{\mathrm{csm}}$ and $R_{\mathrm{csm}}$, the FS1 emission often appears as a pre-peak bump, similar to the breakout signal expected when FS1 breaks out of the ejecta without surrounding CSM. For larger $M_{\mathrm{csm}}$ and $R_{\mathrm{csm}}$, the FS1 emission generally becomes more prominent at peak and persists for a longer time, producing a flatter light-curve profile. However, the duration of the FS1 emission does not increase monotonically with $R_{\mathrm{csm}}$.
For a fixed CSM mass, increasing $R_{\mathrm{csm}}$ does not necessarily delay the peak. In some cases, a larger CSM radius produces an earlier peak than an intermediate value. This behavior arises from the role of the photospheric radius in shaping the light curve.
Specifically, the photospheric radius is given by
\begin{equation}
R_{\rm ph}
\simeq
R_{\rm csm}\left(
1+\frac{M_{\mathrm{thin}}}{M_{\mathrm{csm}}}
\right)^{\!\frac{1}{1-s}},
\end{equation}
where $M_{\mathrm{thin}}=8\pi R_{\mathrm{csm}}^2(s-1
)/[3(3-s)\kappa]$ is a characteristic mass corresponding to an optical depth of $\sim2/3$. In the regime  $M_{\mathrm{csm}} \ll M_{\mathrm{thin}}$ and thus $R_{\mathrm{ph}}  \ll R_{\mathrm{csm}}$,  the light curve enters the cooling phase long before FS1 reaches $R_{\mathrm{csm}}$.
As FS1 continues to move outward between $R_{\mathrm{ph}}$ and $R_{\mathrm{csm}}$, the electron cooling time increases and eventually exceeds the dynamical timescale of the shock. Shock-generated radiation is no longer efficiently thermalized or reprocessed into a quasi-blackbody, and FS1 may evolve adiabatically and generate non-thermal electrons \citep{Suzuki2018}.  These optically thin emission processes are beyond the scope of our calculation.

\subsection{Comparison with Observations}

\begin{figure*}
    \centering
    \includegraphics[width=0.45\linewidth]{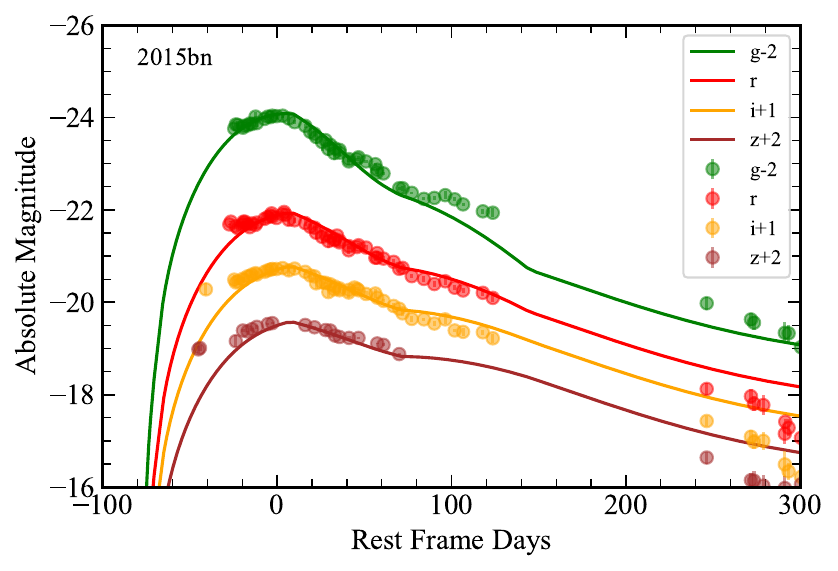}
\includegraphics[width=0.45\linewidth]{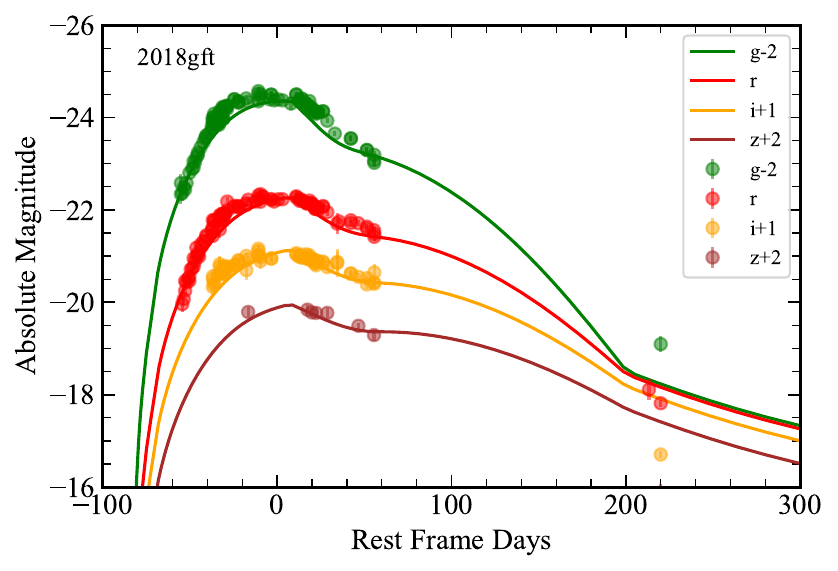}
    \includegraphics[width=0.45\linewidth]{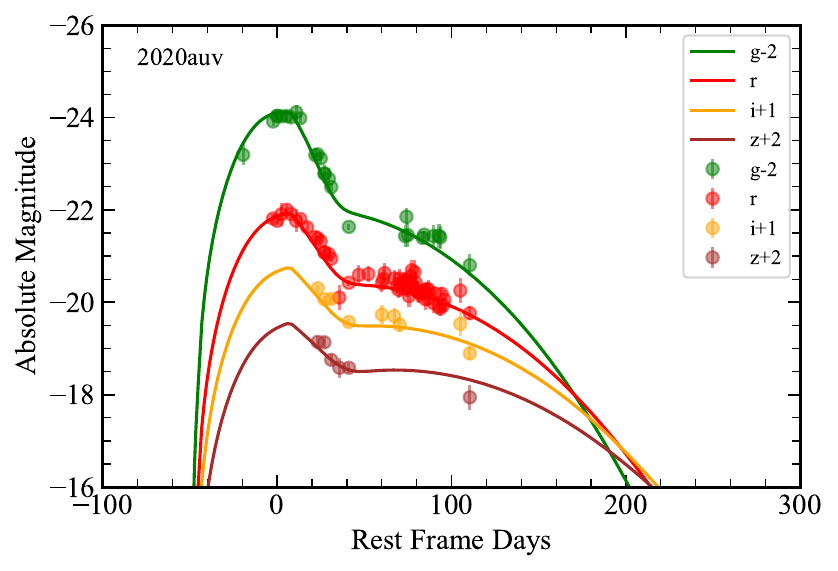}
    \caption{}{Comparison of the multi-band light curves of SN~2015bn, SN~2018gft and SN~2020auv with model light curves, where the data are taken from the Catalog of Type I SLSNe \protect \footnote{\url{https://github.com/gmzsebastian/SLSNe}}~\citep{Gomez2024}.}  
    \label{mlc}
\end{figure*}

\begin{table*}
    \centering
    \caption{Representative parameters used for comparison.}
    \begin{tabular}{ccccccccc}
    \hline
    \hline
    Name & $M_\mathrm{ej}$ & $E_{\mathrm{sn}}$ & $M_{\mathrm{csm}}$ & $R_{\mathrm{csm}}$ & $L_{\rm sd,i}$   & $t_{\rm sd}$   & $\kappa$  & $\kappa_{\gamma}$ \\
    & $M_{\mathrm{\odot}}$ & $\mathrm{erg}$ & $M_{\mathrm{\odot}}$ & $\mathrm{cm}$ & $\mathrm{erg~s^{-1}}$ & $\mathrm{s}$ & $ \mathrm{cm^{2}g^{-1}}$ & $\mathrm{cm^{2}g^{-1}}$
    \\
    \hline
    SN~2015bn  & $ 9 $ & $10^{51}$ & $10$ &
    $4\times10^{15}$ & $3\times10^{45}$ & $2\times10^{6}$ & $0.2$ & $0.01$
    \\
    SN~2018gft  & $10 $ & $10^{51}$ & $8$ &
    $5\times10^{15}$ & $2\times10^{45}$ & $5\times10^{6}$ & $0.2$ & $0.01$
    \\
    SN~2020auv & $12$ & 
    $10^{51}$ & $5 $ & $4\times10^{15}$ & $10^{46}$ & $6\times10^{5}$ & $0.2$ & 0.1 \\
    \hline
    \end{tabular}
    
    \label{para}
\end{table*}
Observationally, SLSNe~I are often classified into fast- and slow-evolving events.  Although the standard magnetar model has sufficient parameter space to reproduce the light curves of both classes, slow-evolving events more frequently show undulations that deviate from the smooth light curves predicted by the magnetar model \citep{Inserra2019}.
Recent population studies have shown markedly asymmetric light-curve evolution in a subclass of SLSNe~I \citep{Chen2023a, Chen2023b}, characterized by longer rise times but more rapid post-peak fading. This behavior is difficult to reconcile with a single-component magnetar model, which generally predicts a more symmetric light-curve morphology \citep{Yu2015}, and favors an interaction scenario. 
Motivated by these considerations, we examine whether a unified hybrid model can broadly reproduce observed SLSN~I light curves.
For illustrative purposes, we apply our model to the nearby event SN~2015bn \citep{Nicholl20162015bn_a, Nicholl20162015bn_b} and two events \citep[ SN~2018gft and SN~2020auv;][]{Chen2023a} from the ZTF sample.
Figure~\ref{mlc} compares the resulting model multi-band light curves with the observations, and the corresponding parameters are listed in Table~\ref{para}. In this comparison, we do not attempt to place stringent constraints on individual parameter values. Instead, we adopt a typical CCSN kinetic energy of $10^{51}~\rm erg$ for the initial explosion, while additional energy is supplied by magnetar spin-down. 
These parameter choices suggest that at least a subset of SLSNe can be understood within the framework of our model, in which roughly half of the injected magnetar energy is converted into the kinetic energy of FS1, helping to meet the high kinetic energy requirement for the interaction scenario. The steep post-peak declines can also arise naturally when the interaction component dominates the peak and is subsequently followed by shock cooling.  We neglect any additional contribution from $^{56}$Ni decay, because producing enough nickel to appreciably affect the SLSN light curve is difficult for ordinary CCSNe. Instead, the late-time emission is powered by magnetar energy stored as internal energy in the PWN.

\subsection{Post-peak bump}

\begin{figure}[htbp]
    \centering
    \includegraphics[width=1\linewidth]{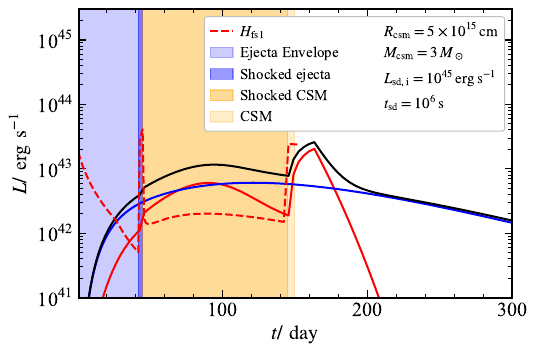}
    \caption{Bolometric luminosity evolution for a parameter set that produces a post-peak bump. Different background colors indicate the region in which FS1 is moving.}
    \label{lightcurve_bump}
\end{figure}

Post-peak bumps are common in SLSNe~I light curves \citep[e.g.,][]{Chen2023a, Hosseinzadeh2022}. Several mechanisms have been proposed for such features, including magnetar flare activity \citep{Yu2017b, Dong2023}, precessing magnetar engines
\citep{Zhang2025,Farah2026}, time-variable thermalization or energy injection \citep{Moriya2022}, and delayed diffusion associated with binary-related evaporated material \citep{Zhu2024}. In our model, a post-peak bump can also arise when FS1 encounters the CSI region at late times and subsequently breaks out of FS2 near $R_{\mathrm{ph}}$. This situation differs from the energetic case with $E_{\mathrm{m}}>E_{\mathrm{sn}}$, in which FS1 encounters the CSI region early and at large optical depth. For $E_{\mathrm{m}}\sim E_{\mathrm{sn}}$ and certain CSM parameters, FS1 moves outward more slowly through the ejecta and reaches the RS at a later time. As shown in Figure~\ref{lightcurve_bump}, FS1 interacts with the CSI region and subsequently breaks out of FS2 near $R_{\mathrm{ph}}$. The strong contrast in density and velocity between the CSI region and the unshocked CSM leads to a sudden transition in the shock heating rate, which can be traced by the emergent radiation and appears as a post-peak bump in the light curve \footnote{The realistic morphology of the post-peak bump is more complex. In our calculations, we adopt a 1D, spherically symmetric, self-similar solution for the CSI region and neglect the impact of the interaction on the CSI structure. However, detailed hydrodynamical simulations indicate that the CSI region has a slightly flatter structure and can be further modified by interacting with FS1 \citep[e.g.,][]{Chevalier1992, Blondin1996,Blondin2001}.}
while the peak is powered by the PWN emission. This scenario requires specific parameters, as discussed in Section~\ref{sec:para}, and is therefore unlikely to explain all post-peak bumps in SLSNe. Rather, the FS1-CSI collision provides a possible channel for producing bump-like features. Other mechanisms, such as engine variability, CSM asymmetry, clumping, or anisotropic energy injection from the central engine, may also contribute.
                
\subsection{Parameter space}\label{sec:para}
Once FS1 breaks out of FS2 and begins to interact with the unshocked CSM, its kinetic energy is efficiently converted into internal energy and subsequently radiated.
To clarify how the dynamical and emission behaviors depend on the magnetar, SN, and CSM parameters, we use the asymptotic solutions discussed in Section~\ref{sec:dyn}
to derive analytical estimates below.
Specifically, given that $\beta<1$ and $\alpha> 1$, the breakout time at which FS1 reaches FS2 can be estimated by setting $R_{\mathrm{fs1}}=R_{\mathrm{fs2}}$ as
\begin{equation}\label{eqs:tb}
   t_{\mathrm{b}}\simeq t_{\mathrm{c}}\begin{cases}
\displaystyle\left[\zeta_{\mathrm{p}}\left(\frac{E_{\mathrm{m}}}{E_{\mathrm{sn}}}\right)^{-1} \left(\frac{t_{\rm sd}}{t_{\mathrm{c}}}\right)\right]^{\lambda},&	 t_{\mathrm{b}}\leq t_{\rm sd},	\\
    \displaystyle
\left[\zeta_{\mathrm{p}}\left(\frac{E_{\mathrm{m}}}{E_{\mathrm{sn}}}\right)^{-1}\right]^{\frac{\lambda}{1-\lambda}},&	t_{\mathrm{b}}>t_{\rm sd},	\\
\end{cases}
\end{equation} 
and the corresponding radius is 
\begin{equation}\label{eqs:Rb}
   R_{\mathrm{b}} \simeq v_{\mathrm{t}}t_{\mathrm{c}} \begin{cases}
\displaystyle\left[\zeta_{\mathrm{p}}\left(\frac{E_{\mathrm{m}}}{E_{\mathrm{sn}}}\right)^{-1}
\left(\frac{t_{\rm sd}}{t_{\mathrm{c}}}\right)\right]^{\beta\lambda}, &t_{\mathrm{b}}\leq t_{\rm sd},\\ \displaystyle
\left[\zeta_{\mathrm{p}}\left(\frac{E_{\mathrm{m}}}{E_{\mathrm{sn}}}\right)^{-1}\right]^{\frac{\beta\lambda}{1-\lambda}},&t_{\mathrm{b}}> t_{\rm sd},
    \end{cases}
\end{equation}
where $\lambda=(\alpha-1)/(\alpha-\beta)$.
By comparing $t_{\mathrm{c}}$, $t_{\mathrm{p}}$, and $t_{\mathrm{b}}$, the conditions for the breakout to occur in the outer envelope of the ejecta can be written as
\begin{equation}\label{tinout}
  \frac{t_{\rm sd}}{t_{\mathrm{c}}}\lesssim\zeta_{\mathrm{p}}^{-1}\frac{E_{\mathrm{m}}}{E_{\mathrm{sn}} }\;\mathrm{and}\;\frac{E_{\mathrm{m}}}{E_{\mathrm{sn}} }\gtrsim \zeta_{\mathrm{p}}.
\end{equation}
The condition for the breakout to occur before $t_{\rm sd}$ is  
\begin{equation}\label{tcom}
    \frac{t_{\rm sd}}{t_{\mathrm{c}}}\gtrsim\left(\frac{E_{\mathrm{m}}}{E_{\mathrm{sn}}}\right)^{\frac{\lambda}{\lambda-1}}\zeta_{\mathrm{p}}^{\frac{\lambda}{1-\lambda}}.
\end{equation}

\begin{figure}[htbp]
    \centering
    \includegraphics[width=1\linewidth]{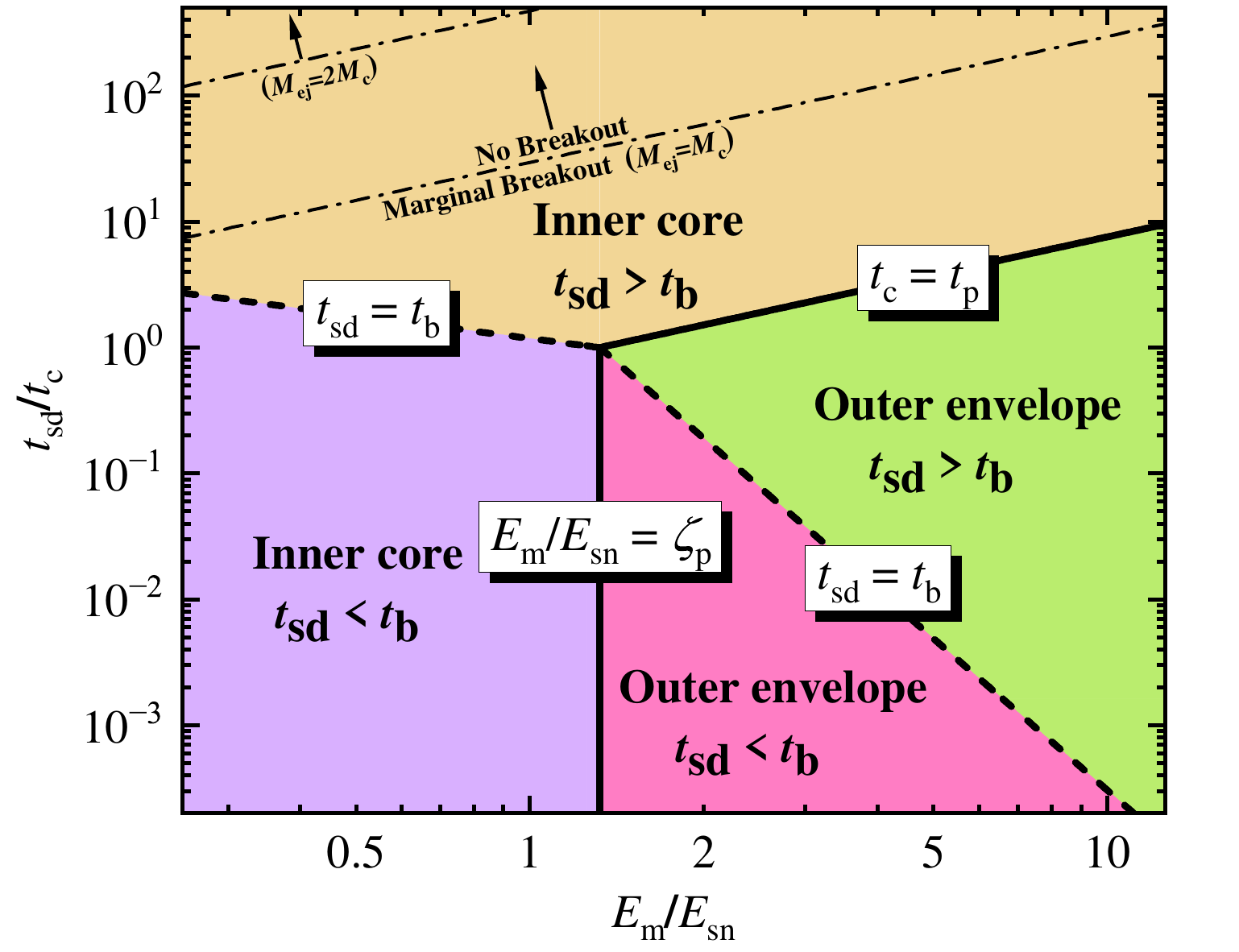}
    \caption{Different breakout regions of  FS1 from FS2, illustrating whether the breakout occurs in the inner core or outer envelope of the SN ejecta and whether the magnetar has spun down at the time of breakout. The region above the dashed-dot line implies that FS1 cannot break out of FS2 before FS2 enters the optically thin region. }
    \label{csmpwn}
\end{figure}

Furthermore, if the breakout occurs near the photospheric radius (i.e., $R_{\mathrm{b}} \simeq R_{\mathrm{ph}}$), the following condition should be satisfied:
\begin{equation}\label{eqs:MejMc}
\frac{M_{\mathrm{ej}}}{M_{\mathrm{c}} }\simeq
\zeta_{\mathrm c}^{s-3}
\begin{cases}
\displaystyle
\left[\zeta_{\mathrm p}\left(\frac{E_{\mathrm m}}{E_{\mathrm{sn}}}\right)^{-1}
\left(\frac{t_{\rm sd}}{t_{\mathrm c}}\right)\right]^{(s-3)\beta\lambda},
& t_{\mathrm b}\le t_{\rm sd},\\
\displaystyle
\left[\zeta_{\mathrm p}
\left(\frac{E_{\mathrm m}}{E_{\mathrm{sn}}}\right)^{-1}\right]^{\frac{s-3}{1-\lambda}\beta\lambda},
& t_{\mathrm b} > t_{\rm sd},
\end{cases}
\end{equation}
where $M_{\mathrm{c}}= M_{\mathrm{csm}}(R_{\mathrm{ph}}/R_{\mathrm{csm}})^{3-s} =M_{\mathrm{csm}}(1+M_{\mathrm{thin}}/M_{\mathrm{csm}})^{(s-3)/(s-1)}$ is the CSM mass enclosed within the photospheric radius.
Equations~(\ref{tinout}) and ~(\ref{tcom}) separate the $ E_{\mathrm{m}}/E_{\mathrm{sn}}-t_{\rm sd}/t_{\mathrm{c}}$ plane into four distinct regions as illustrated in Figure~\ref{csmpwn}. 

This partition shows how the breakout behavior depends jointly on the magnetar, ejecta, and CSM parameters.  It indicates whether FS1 catches up with FS2 in the inner core or in the outer envelope of the ejecta, and whether the breakout occurs before or after the magnetar spin-down timescale. For the energetic magnetars considered here, the most relevant case corresponds to collision and breakout in the outer envelope. In the opposite region of parameter space, the collision occurs deeper in the ejecta, or FS1 may fail to catch up with the RS before the CSI region becomes optically thin.
In addition to the location and timing of the breakout,  another constraint comes from the dynamics of FS1 as it interacts with the CSI region. The pressure exerted on FS1 when it interacts with the CSI region can be estimated as $P_{\mathrm{csi}}\sim P_{\mathrm{fs2}}\sim \rho_{\mathrm{csm}} v_{\mathrm{fs2}}^2 $, while the pressure within the PWB  can be estimated as
\begin{equation}
    P_{\mathrm{pwb}}\sim \begin{cases}
        \displaystyle\frac{E_{\mathrm{m}}t}{4\pi R_{\mathrm{fs2}}^3 t_{\rm sd}}, &t\leq t_{\rm sd},  \\
            \displaystyle\frac{E_{\mathrm{m}}}{4\pi R_{\mathrm{fs2}}^3}, & t> t_{\rm sd}.
    \end{cases}
\end{equation}
The ratio of these two pressures at $t_{\mathrm{b}}$ is
\begin{equation}
\frac{P_{\mathrm{pwb}}}{P_{\mathrm{csi}}}\sim
\frac{\zeta_{\mathrm c}^{s-3}\zeta_v^{-2}\zeta_{\mathrm{p}}}{3-s}
\begin{cases}
\displaystyle
\left[\zeta_{\mathrm p}\left(\frac{E_{\mathrm m}}{E_{\mathrm{sn}}}\right)^{-1}
\left(\frac{t_{\rm sd}}{t_{\mathrm c}}\right)\right]^{k_1\lambda-1},
& t_{\mathrm b}\le t_{\rm sd},\\
\displaystyle
\left[\zeta_{\mathrm p}
\left(\frac{E_{\mathrm m}}{E_{\mathrm{sn}}}\right)^{-1}\right]^{\frac{k_2\lambda}{1-\lambda}-1},
& t_{\mathrm b} > t_{\rm sd},
\end{cases}
\end{equation}
where $k_1= 3+(s-5)\beta$ and $ k_2= 2+(s-5)\beta$.
\begin{figure*}[htbp]
    \centering
\includegraphics[width=1.0\linewidth]{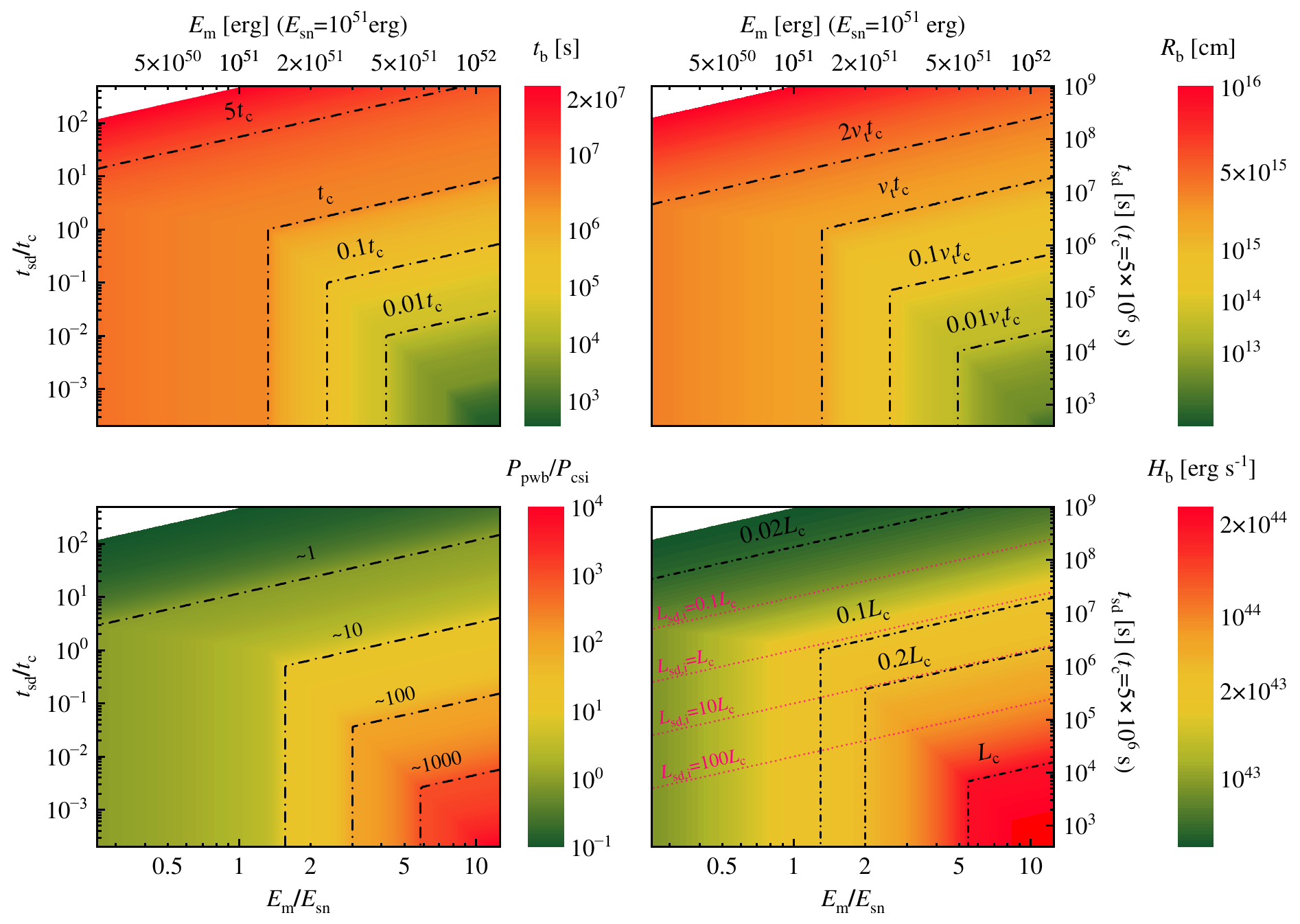}
    \caption{Variations of the breakout time $t_{\mathrm{b}}$, radius $R_{\mathrm{b}}$, $P_{\mathrm{pwb}}/P_{\mathrm{csi}}$, and the heating rate $H_{\mathrm{b}}$ in the $E_{\mathrm{m}}/E_{\mathrm{sn}}-t_{\mathrm{sd}}/t_{\mathrm{c}}$ parameter space. The left and bottom axes show the plane spanned by $ E_{\mathrm{m}}/E_{\mathrm{sn}}$ and $t_{\rm sd}/t_{\mathrm{c}}$, while the top and right axes indicate the corresponding values of $E_{\mathrm{sn}}$ and $t_{\rm c}$ obtained for the representative choices.}
    \label{copara}
\end{figure*}

The peak luminosity of the FS1 emission can be roughly estimated from its heating rate. When FS1 interacts with the CSM, the heating rate is
\begin{equation}
    H_{\mathrm{fs1}}\simeq2\pi R_{\mathrm{fs1}}^2\rho_{\mathrm{csm}}v_{\mathrm{fs1}}^3\simeq \frac{(3-s) M_{\mathrm{csm}}}{2R_{\mathrm{csm}}^{3-s}}R_{\mathrm{fs1}}^{\mathrm{2-s}}v_{\mathrm{fs1}}^3.
\end{equation}
FS1 is strongly accelerated before $t_{\mathrm{sd}}$ and subsequently follows an almost ballistic trajectory as it interacts with the CSM.  The shock velocity $v_{\mathrm{fs1}}$ remains close to the breakout velocity $ v_{\mathrm{b}}$ until substantial CSM is swept up. $H_{\mathrm{fs1}}$ can be estimated by its value at breakout as
\begin{equation}\label{eqs:Hb} 
    \begin{aligned}
            &H_{\mathrm{fs1}}\simeq H_{\mathrm{b}}\simeq\\
            &
    \frac{(3-s)}{2}\zeta_{v}^2\zeta_{\mathrm{c}}^{3-s}L_{\mathrm
            c}
    \begin{cases}
     \displaystyle
        \left[\zeta_{\mathrm{p}}
        \left(\frac{E_{\mathrm{m}}}{E_{\mathrm{sn}}}\right)^{-1}\left(\frac{t_{\rm sd}}{t_{\mathrm{c}}}\right) \right]^{k_3\lambda}, & 
    
        t_{\mathrm{b}}\leq t_{\rm sd}, \\
         \displaystyle
        \left[\zeta_{\mathrm{p}}\left(\frac{E_{\mathrm{m}}}{E_{\mathrm{sn}}}\right)^{-1}\right]^{\frac{k_3\lambda }{1-\lambda}}, &  t_{\mathrm{b}}>t_{\rm sd},
    \end{cases}
    \end{aligned}
\end{equation}
where $k_3=\beta(5-s)-3$ and $L_{\mathrm{c}}=E_{\mathrm{sn}}/t_{\mathrm{c}}$ is a characteristic luminosity.
Figure~\ref{copara} summarizes how  $t_{\mathrm{b}}$, $R_{\mathrm{b}}$, $P_{\mathrm{pwb}}/P_{\mathrm{csi}}$ and $H_{\mathrm{b}}$ vary across the parameter space.
As shown in the figure, releasing a large amount of rotational energy from the magnetar over a short spin-down timescale produces a more energetic FS1. In particular, a higher ratio of $L_{\mathrm{sd,i}}/L_{\mathrm{c}}$ corresponds to a smaller breakout time $t_{\mathrm{b}}$ and radius $R_{\mathrm{b}}$.
Nevertheless, although $L_{\rm sd,i}$ increases steeply with increasing $E_{\mathrm{m}}$ and decreasing $t_{\mathrm{sd}}$, the corresponding heating rate $H_{\mathrm{b}}$ depends more weakly on these parameters. This
indicates that the detailed energy injection history has a limited impact on the dynamics of FS1. The high internal pressure of the PWB (with $P_{\mathrm{pwb}} \gg P_{\mathrm{csi}}$) ensures that FS1 is only mildly affected by the CSI region and can successfully break out of FS2.  
Conversely, in the upper-left part of the parameter space, where $t_{\rm sd}$ is relatively large and $L_{\mathrm{sd,i}}$ is small, FS1 encounters the inward-moving RS before being efficiently accelerated. In this case, FS1 can be decelerated or even compressed by the pressure exerted by the CSI region, roughly corresponding to $P_{\mathrm{pwb}} \lesssim P_{\mathrm{csi}}$. 

If FS1 is not strongly decelerated or compressed by the CSI region and breaks out of FS2 near $R_{\mathrm{b}}\simeq R_{\mathrm{ph}}$, it can produce a post-peak bump. The characteristic time and luminosity of this bump can be derived by substituting Equation~(\ref{eqs:MejMc}) into Equation~(\ref{eqs:tb}) and Equation~(\ref{eqs:Hb}):
\begin{equation}\label{eqs:tpb}
t_\mathrm{pb}\simeq t_\mathrm{c}\left[\zeta_\mathrm{c}^{s-3}\left(\frac{M_\mathrm{c}}{M_{\rm ej}}\right)\right]^{\frac{1}{(3-s)\beta}}
\end{equation}
and
\begin{equation}\label{eqs:Lpb}
L_{\mathrm{pb}}\simeq H_\mathrm{pb} \simeq \frac{3-s}{2}\zeta_v^2\zeta_\mathrm{c}^{3-s}
\left[\zeta_\mathrm{c}^{3-s}\left(\frac{M_{\rm ej}}{M_\mathrm{c}}\right)\right]^{\frac{k_3}{(s-3)\beta}}L_\mathrm{c}.
\end{equation}
The observational properties of the post-peak bump can then be used to infer the SN and CSM parameters as
\begin{equation}
M_{\rm ej}\simeq \zeta_\mathrm{c}^{s-3} 
\left(\frac{\zeta_{\mathrm{pb}}E_{\rm sn}}{ L_\mathrm{pb}t_\mathrm{pb}}\right)^{k_4}
\begin{cases}
\displaystyle
M_{\rm csm},   \\\quad\quad\quad\quad R_{\mathrm{ph}}\sim R_{\rm csm},
\\
\displaystyle
\left[\frac{16\pi(s-1)}{3(3-s)^2\kappa}\right]^{1/2}
t_{\mathrm{pb}}^{3/2}L_{\mathrm{pb}}^{1/2},  \\\quad\quad\quad\quad R_{\rm ph}\ll R_{\rm csm},
\end{cases}
\end{equation}
where $\zeta_{\mathrm{pb}}=(3-s)\zeta_\mathrm{c}^{3-s}\zeta_v^2/2$ and $k_4=(3-s)\beta/[\beta(5-s)-2]$. 
These inferences are valid only when the post-peak bump occurs while the RS is still in the outer ejecta, i.e. $t_{\rm pb}/t_{\rm c} <1$. Combining Equation~(\ref{eqs:tpb}) and Equation~(\ref{eqs:Lpb}) gives $L_{\rm pb}t_{\rm pb}/\zeta_{\rm pb}E_{\rm sn}=(t_{\rm pb}/t_{\rm c})^{k_3+1}$. For $t_{\rm pb}/t_{\rm c} >1$, the CSI region approaches the Sedov-Taylor solution with $\beta=2/(5-s)$ and hence $k_3+1=0$. In this limit, $L_{\rm pb}t_{\rm pb}\simeq \zeta_{\rm pb}E_{\rm sn}$ and the mass constraints become degenerate, so the post-peak bump time and luminosity provide only an estimate of the kinetic energy scale rather than a constraint on $M_{\rm ej}$.

\section{Summary and discussion}\label{Sec: Sum}
In this paper, we develop a semi-analytical hybrid model to describe the dynamical evolution of a PWB driven by a millisecond magnetar as it interacts with both the SN ejecta and the surrounding CSM. In contrast to previous approaches in which the magnetar energy injection and CSI are treated as two independent components, our model follows the interactions among the PWB, the SN ejecta, and the CSM. These interactions naturally give rise to multi-component emission, with different components dominating at different epochs or contributing simultaneously. Within this framework, we show that the additional energy required by interaction scenarios can in principle be supplied by magnetar injection on top of an otherwise canonical core-collapse SN, without invoking an unusually energetic explosion. 
We also investigate how the partition of magnetar input among bulk acceleration, shock heating and radiation depends on the underlying magnetar, ejecta, and CSM parameters.  
Accordingly, the hybrid model can produce a range of light-curve behaviors, including different rise times, peak luminosities, and post-peak decline rates. Under some conditions, the interaction between FS1 and the CSI region can also introduce bump-like features in the light curve. 
These results suggest that at least some of the observed diversity of SLSNe can be understood within our hybrid model. 

The dense CSM required in interaction scenarios likely reflects enhanced mass loss shortly before core collapse \citep{Smith2014}. Several channels, besides PPI ejections, may contribute to the formation of such a CSM, including wave-driven envelope mass loss \citep{Quataert2012, Shiode2014}, super-Eddington outflows \citep{Smith2006, Quataert2016}, and binary interaction \citep{Chevalier2012b, Sana2012, Ivanova2013}. In the specific context considered here, a binary channel can be particularly attractive. The formation of a rapidly rotating magnetar requires the progenitor to retain sufficient angular momentum until core collapse. Binary interaction provides a plausible way to retain such rapid rotation through tidal spin-up or angular-momentum transfer from a companion, while mass transfer or common-envelope ejection from binary systems could produce dense CSM. This possibility is consistent with the view that binary evolution plays an important role in producing stripped-envelope SN progenitors, since their relatively high event rate and low ejecta masses are difficult to reconcile with rare very massive single-star channels alone \citep{Smith2011, Shivvers2017, Yoon2017, Prentice2019}. 
In such a picture, ordinary stripped-envelope SNe may represent the more common outcome of binary stripping, whereas SLSNe~I could correspond to a rarer subset in which the tightest systems both enhance pre-SN mass loss and retain sufficient core angular momentum to form rapidly rotating magnetars.
Detailed binary-interaction calculations indeed suggest that some stripped systems may explode during ongoing hydrogen-free mass transfer and very close binaries can, in at least some cases, form millisecond magnetars \citep{Hu2023, Ercolino2025}.
This possibility may also help interpret the reported anti-correlation between the inferred magnetar rotational energy and the SLSN ejecta mass \citep{Yu2017, Liu2022, Zhu2026}.  

At the same time, a binary origin also implies that the CSM is unlikely to be strictly spherical. Mass loss via Roche-lobe overflow or common-envelope ejection is expected to be concentrated toward the orbital plane, generating torus- or disk-like dense structures embedded in a more dilute wind. 
However, distinguishing among different progenitor channels and investigating the detailed CSM structure are beyond the scope of the present work. For simplicity,  we approximate the CSM by a smooth, spherically symmetric $\rho_{\rm csm} \propto r^{-2}$ profile. In reality, departures from spherical symmetry and temporal variations in the mass-loss history may introduce viewing-angle effects and small-scale structure in the observed light curves. If the emitting region remains optically thick, however, radiative diffusion can smooth part of this complexity, so that a spherical treatment can still capture the main emission components and characteristic timescales of the multi-component light-curve evolution.

Spectroscopic signatures provide an important complement to light-curve modeling. The currently available spectra of well-observed SLSNe~I around peak have not revealed unambiguous narrow-line signatures of CSI. This, however, does not necessarily rule out interaction. If the CSM is highly ionized and/or the interaction region is embedded within optically thick CSM, energy dissipated by the interaction can be thermalized and reprocessed into a smooth continuum, leaving only subtle spectroscopic signatures. At later phases, some events exhibit broad H$\alpha$ emission lines \citep{Yan2015, Yan2017, Quimby2018}, suggesting high-velocity interaction with hydrogen-rich CSM.  In our model, if the CSM is sufficiently extended, FS1 does not immediately break out of the CSM. Instead, it continues to move into optically thin outer CSM, which could include hydrogen-rich material lost during earlier evolutionary stages. In addition, different from a pure CSI scenario, in which the shock is expected to decelerate as swept-up mass increases, continued energy injection from the magnetar can instead help maintain a high FS1 velocity, offering a possible explanation for the high-velocity broad-line features seen in some SLSNe~I \citep[e.g.,][]{Liu2017Spec, Quimby2018, Aamer2025}. The velocity evolution of such late-time broad lines may therefore provide useful diagnostics of our hybrid model.

\end{CJK*}
\section*{Acknowledgments}
This work is supported by the National Natural Science Foundation of China (grant Nos. 12393811 and 12303047), the National SKA Program of China (2020SKA0120300), and the China Manned Spaced Project (CMS-CSST-2021-A12).



\bibliography{sample631}{}
\bibliographystyle{aasjournal}

\end{document}